\documentclass[11pt,twoside]{article}
\usepackage{asp2010}
\usepackage{caption}
\begin{document}

\title{Updates on the Pulsating SdB Star Feige 48 through Spectroscopy}
\author{M. Latour$^1$, G. Fontaine$^1$, and E. M. Green$^2$
\affil{$^1$D\'epartement de Physique, Universit\'e
  de Montr\'eal, Succ. Centre-Ville, C.P. 6128, Montr\'eal, QC H3C 3J7,
  Canada}
\affil{$^2$Steward Observatory, University of Arizona, 933 North
  Cherry Avenue, Tucson, AZ 85721 }}

\begin{abstract}
As part of a concentrated effort to understand and exploit better the seismic properties of Feige 48, we reestimated its atmospheric parameters using, for the first time, a self-consistent approach based on the detailed metal abundances that have been determined in previous studies of both its UV and optical spectrum. We computed a small 3D grid of 150 fully line-blanketed, NLTE model atmospheres incorporating the 8 most abundant metals found in the atmosphere of that star. We next compared quantitatively the synthetic spectra that we calculated from these models with three observed spectra having different resolutions (1 \AA, 6 \AA, and 8.7 \AA) and wavelength ranges. Our findings are $T_{\rm eff}$ = 29,890 K, log $g$ = 5.46, and log $N$(He)/$N$(H) =  $-$2.88, with very small formal fitting errors. These estimates are in good agreement with those of previous studies, thus showing that the presence of metals do not affect significantly the bulk atmospheric properties of Feige 48. We also modeled the He \textsc{ii} line at 1640 \AA~from the $STIS$ spectrum of the star and we found with this line an effective temperature and a surface gravity that match well the values obtained with the optical data.

We also present a radial velocity curve based on several years of observing Feige 48 at the MMT. This curve indicates a precise orbital period of 8.25 h. Coupled to our discovery of a weak but very real periodicity of 8.25 h in the light curve of Feige 48, that was gathered during a recent extensive photometric campaign at the Mount Bigelow 1.55 m Kuiper Telescope, we conclude that there is a reflection effect in the system, and that the unseen companion is a low mass MS star and not a white dwarf as has been previously suggested. 
\end{abstract}

\section{Introduction}
Feige 48 is a well known rapidly pulsating subdwarf B (sdB) star. A first asteroseismic analysis of this star has been carried out by \citet{char05} who found the star to have a mass of 0.46 $M_{\rm \odot}$, which is very near the mean mass of the sdB's (0.47 $M_{\rm \odot}$) derived by \citet{font12}. Synchronous rotation between Feige 48 and its invisible companion (until now thought to be a white dwarf), forming a binary system with an orbital period around 9.0 h, has also been established using asteroseismology \citep{vang08}. Their orbital period was in agreement with the one found by \citet{otoole06} from radial velocity variation. 
Both of these asteroseismic results rely on independent estimates of the effective temperature and surface gravity provided by, for example, spectroscopy. This is because degeneracies exist in the seismic solutions that reproduce the observed pulsation periods of Feige 48. It is therefore crucial to obtain the most reliable estimates of the atmospheric parameters that spectroscopy can offer.
%Because degeneracies exist in the seismic solutions that reproduce the observed pulsation periods of Feige 48, it is crucial to obtain the most reliable estimates of the atmospheric parameters that spectroscopy can offer.

\section{Atmospheric Parameters for Feige 48}
In order to fix a suitable chemical composition for our grid of model atmospheres, we relied on four abundance studies that have been done for Feige 48: \citet{heb00} (Keck/$HIRES$), \citet{chay04} ($FUSE$), \citet{otoole06} ($HST/STIS$) and more recently \citet{geier13} (Keck/$HIRES$). We calculated a weighted average of the abundances of each element using these studies and included in our models the eight most abundant metals. Our chemical composition is detailed in Table 1.

We computed our NLTE line-blanketed model atmospheres using the TLUSTY and SYNSPEC codes, and made a small grid of 150 models especially suited for Feige 48. We used this grid to fit three different spectra of Feige~48, they cover different optical wavelength ranges and resolutions (1 \AA, 6 \AA, and 8.7 \AA). The resulting fit for the 1~\AA~MMT combined spectrum (see section 3) is shown in Figure 1. The weighted mean parameters we obtained with the three spectra are $T_{\rm eff}$ = 29,890 K, log $g$ = 5.46, and log $N$(He)/$N$(H) =  $-$2.88. These parameters are similar to what has been determined in the past by different studies (\cite{koen98}, \cite{heb00} and \cite{char05}) using different type of models (LTE, NLTE, with and without metals).

Because of its relatively cool temperature and its low helium abundance, no He~\textsc{ii} lines are visible in the optical spectra, there was not even a hint of He~\textsc{ii} $\lambda$4686 in the HIRES spectrum of \citet{heb00}. Since the star has been observed in the UV range with the $STIS$ spectrograph, we retrieved the data and checked for the He~\textsc{ii} 1640~\AA~line. We indeed found something corresponding to this line and tried to fit it with our model grid. We were able to converge to a unique solution by fixing the helium abundance and letting the program find the optimal temperature and gravity. As seen in Figure 2, the values found with the 1640 \AA~line correspond well (within the formal uncertainties of the fitting procedure) with the ones we obtained using the optical spectra.

\begin{minipage}{.55\textwidth}%\centering
\vspace{0.8cm}
\includegraphics[width=13pc,angle=270]{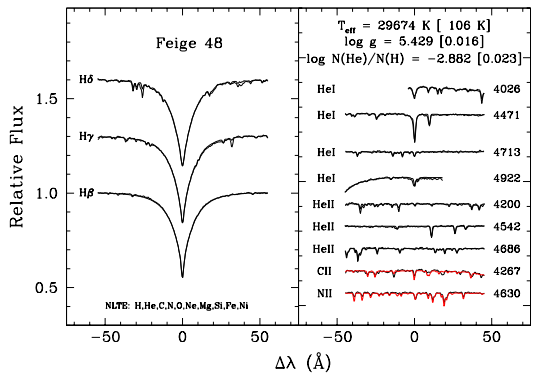}
\captionof{figure}{Resulting fit and parameters for Feige~48}
%\end{center}
\end{minipage}
\begin{minipage}{.4\textwidth}%\centering
%\begin{table}
\captionof{table}{Chemical composition of our models for Feige~48}
\begin{tabular}{ccc}
\hline\noalign{\smallskip}
Elements &  Mean  &  Solar \\
   & \multicolumn{2}{c}{log $N$(X)/$N$(H)}  \\
   \hline
C  & $-$4.65 $\pm$ 0.03 & $-$3.57  \\
N  & $-$4.47 $\pm$ 0.07 & $-$4.17  \\
O  & $-$4.25 $\pm$ 0.10 & $-$3.31  \\
Ne & $-$4.90 $\pm$ 0.31 & $-$4.07 \\
Mg & $-$5.15 $\pm$ 0.35 & $-$4.40 \\
Si & $-$5.49 $\pm$ 0.06 & $-$4.49 \\
Fe & $-$4.41 $\pm$ 0.10 & $-$4.50 \\
Ni & $-$5.31 $\pm$ 0.15 & $-$5.78 \\
\hline\noalign{\smallskip}
\end{tabular}
\end{minipage}

\begin{figure}[t]
\begin{center}
\includegraphics[width=20pc,angle=270]{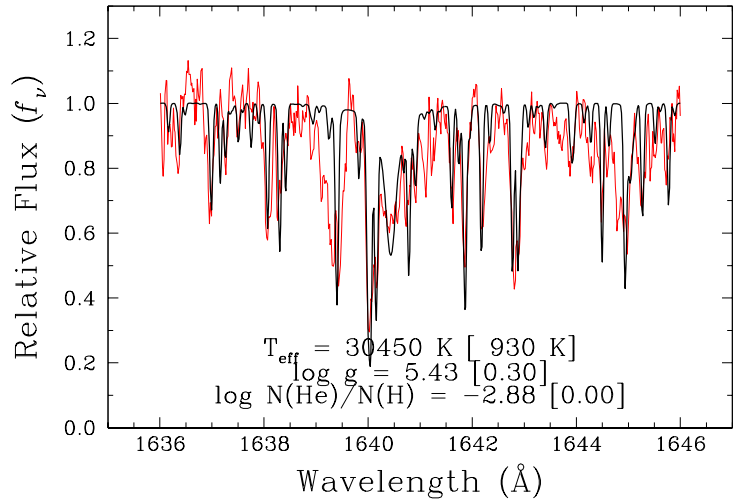}
\caption{Our fit to the He~\textsc{ii} $\lambda$1640 line in the $STIS$ spectrum of Feige 48. The He~\textsc{ii} line is blended with Fe~\textsc{iv} and Ni~\textsc{iii} lines around 1640 \AA~and other Fe~\textsc{iv} lines on its red wing.  }
\end{center}
\end{figure}

\section{Radial Velocity Curve \& Orbital Period}

Eighteen spectra of Feige 48 were acquired with the 6.5 m MMT Blue Spectrograph between 2002 and 2013, one to four spectra per night, using an identical spectroscopic set up for each run. The 832/mm grating in second order and 1$\arcsec$ slit provides a resolution, R, $\sim$4250
(1.05 \AA) over the wavelength range 4000-4950 \AA. Exposures of 240 to 475 s, depending on conditions, resulted in $S/N$ sufficient to achieve velocity errors of 1 to 2 km/s for sdB stars.
After removal of the continuum, the spectra were cross-correlated to derive radial velocities using the IRAF FXCOR routine. 
The Fourier filtering parameters for the cross-correlations were optimized for highest sensitivity to narrow helium and metal lines and the cores of the Balmer lines.
A least-squares period search from 0.1 to 30 days gave a best solution of P = 0.3436086 $\pm$ 0.0000005 d (8.24662 h) with amplitude K = $v sini$ = 25.1 $\pm$ 1.4 km/s. Figure 3 (left panel) shows the corresponding velocity curve and MMT data points vs orbital phase. 
Since the period is so close to 1/3 day, there are still aliases remaining near 1/5, 1/4, 1/2, and 1 day, though they are much less likely.

% \begin{minipage}{.5\textwidth}%\centering
% %\vspace{0.8cm}
% \includegraphics[width=13pc,angle=0]{f48_mmtrv.eps}
% \captionof{figure}{ Velocity curve of Feige 48's MMT spectra vs orbital phase}
% %\end{center}
% \end{minipage}
% \begin{minipage}{.5\textwidth}%\centering
% %\begin{table}
% \includegraphics[width=13pc,angle=0]{grfold4F.eps}
% \captionof{figure}{The result of folding the light curve over the period and binning the orbital phase}
% \end{minipage}
% %\end{table}

\begin{figure}[t!]
\epsscale{1.5}
\plottwo{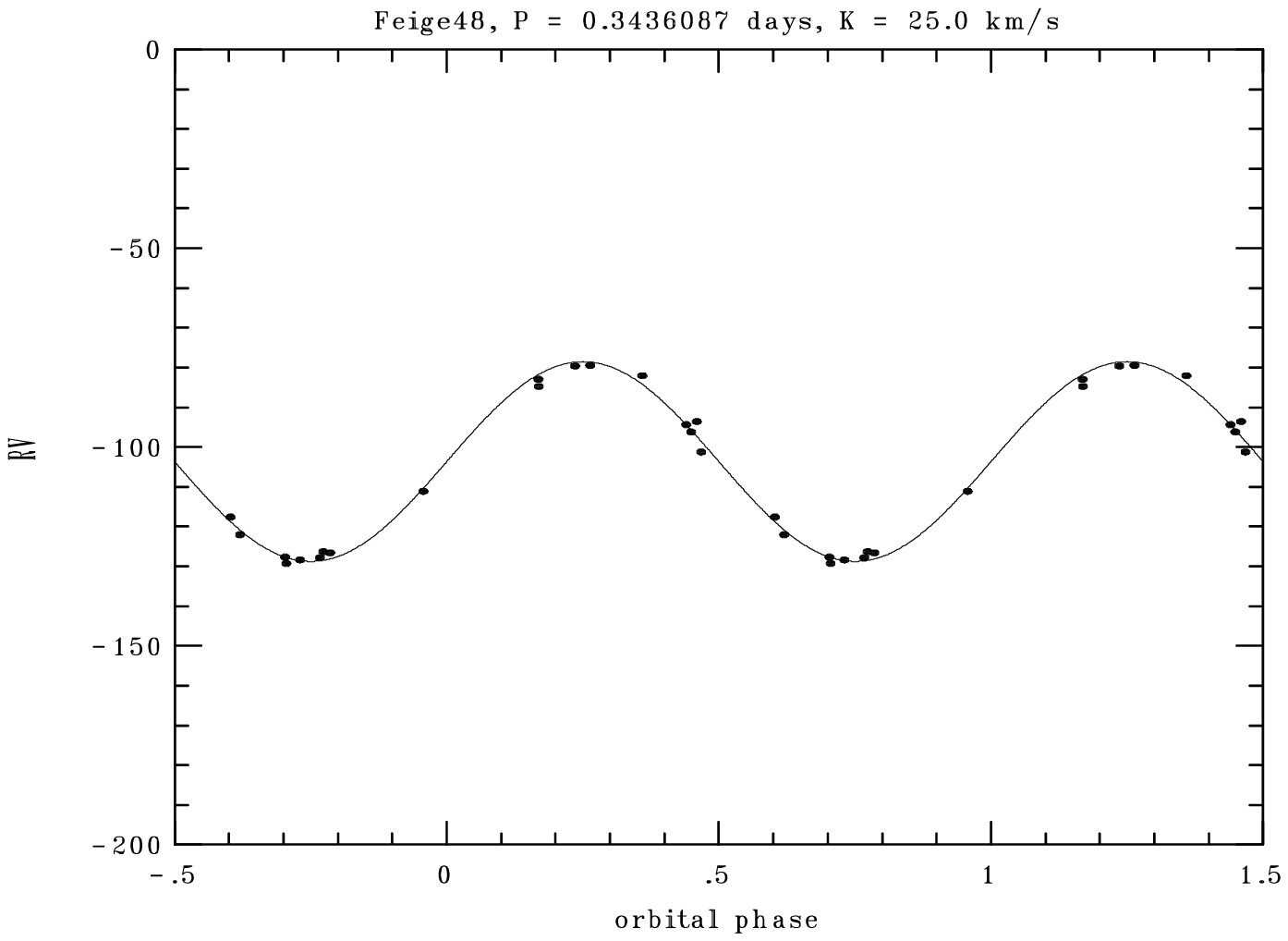}{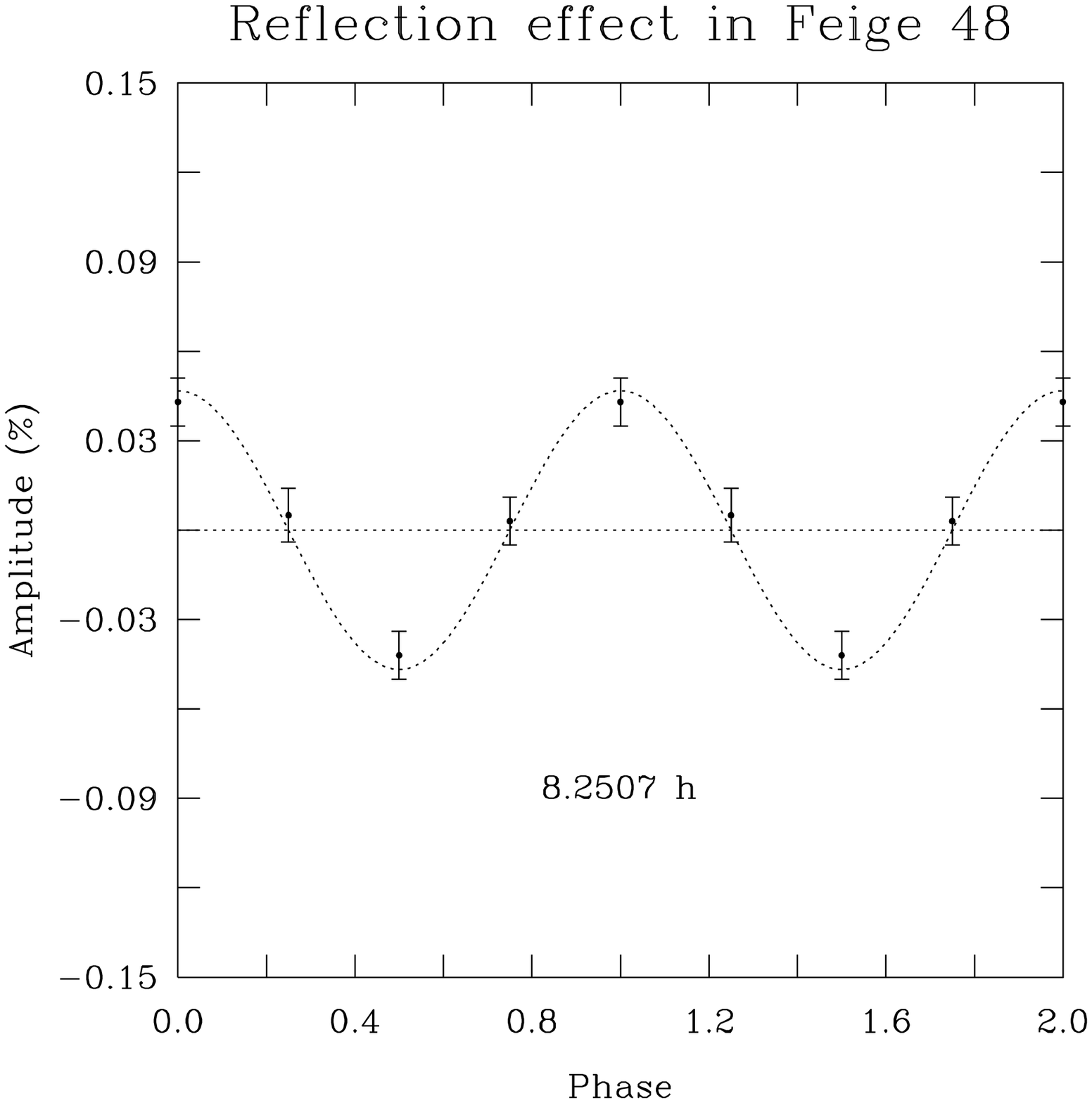}
\caption{Velocity curve of Feige 48's MMT spectra vs orbital phase (Left). The result of folding the light curve over the period and binning the orbital phase (Right).}
\end{figure}

\section{Light Curve \& Companion}

Feige 48 has been extensively observed between January and May 2009 during a photometric campaign with the Mont4K detector at the Mount Bigelow 1.55 m Kuiper Telescope (see Fontaine et al. 2013, these proceedings). Almost 400 h of data were gathered, corresponding to a duty cycle of 12.2\%. 
When looking at the Fourier transform of the light curve, a peak was found at the corresponding period of the orbital system, 8.2507 h. The peak is weak but above the 4 $\sigma$ level. 
Figure 3 (right panel) illustrates the result of folding the light curve over the period and binning the orbital phase.
The sinusoidal curve typical of a reflection effect is clearly seen and thus we can claim that Feige 48's companion is a cool M dwarf because a tiny white dwarf could not produce such a reflection. The amplitude of the luminosity variation is around 0.04\%, corresponding to $\sim$0.01 mag, which is below the upper limit set by \citet{otoole04} who favored at the time the option of a white dwarf companion.

%\acknowledgements 

\bibliography{MlatourF48ref}

\end{document}